\documentclass{article}
\usepackage[preprint,nonatbib]{neurips_2022}
\usepackage{amsmath}
\usepackage{graphicx}
\usepackage{float}
\usepackage{booktabs}
\usepackage{breqn}
\usepackage{makecell}
\usepackage[utf8]{inputenc} 
\usepackage[T1]{fontenc}    
\usepackage{hyperref}       
\usepackage{url}            
\usepackage{booktabs}       
\usepackage{amsfonts}       
\usepackage{nicefrac}       
\usepackage{microtype}      
\usepackage{xcolor}         
\usepackage{url} 
 \usepackage{caption}
\usepackage{subcaption}

\usepackage{xcolor}

  \title{Smoothing spline analysis of variance models: A new tool for the analysis of accelerometer data}
\author{ 
  Rui Xie\thanks{Authors contributed equally. This work is partially supported by NIH awards R01MD018025 and R03AG069799.} \\
  Department of Statistics and Data Science\\
  University of Central Florida\\
  \And
  Lulu Chen$^\ast$\\Department of Statistics and Data Science\\
  University of Central Florida\\
  \And
  Joon-Hyuk Park\\
  Department of Mechanical and Aerospace Engineering \\
  University of Central Florida \\
  \And
  Jeffrey Stout \\
  School of Kinesiology and Rehabilitation Sciences \\
  University of Central Florida \\
  \And
  Ladda Thiamwong\\
  College of Nursing \\
  University of Central Florida \\
}

\begin{document}

\maketitle
\begin{abstract}
Accelerometer data is commonplace in physical activity research, exercise science, and public health studies, where the goal is to understand and compare physical activity differences between groups and/or subject populations, and to identify patterns and trends in physical activity behavior to inform interventions for improving public health.
We propose using mixed-effects smoothing spline analysis of variance (SSANOVA) as a new tool for analyzing accelerometer data.
By representing data as functions or curves, smoothing spline allows for accurate modeling of the underlying physical activity patterns throughout the day, especially when the accelerometer data is continuous and sampled at high frequency.
The SSANOVA framework makes it possible to decompose the estimated function into the portion that is common across groups (i.e., the average activity) and the portion that differs across groups. By decomposing the function of physical activity measurements in such a manner, we can estimate group differences and identify the regions of difference. In this study, we demonstrate the advantages of utilizing SSANOVA models to analyze accelerometer-based physical activity data collected from community-dwelling older adults across various fall risk categories. 
Using Bayesian confidence intervals, the SSANOVA results can be used to reliably quantify physical activity differences between fall risk groups and identify the time regions that differ throughout the day.
\end{abstract}

\noindent%

\section{Introduction}
Accelerometer data collected from wearable devices for monitoring daily physical activity is commonly used in physical activity research, exercise science, and fall risk intervention studies\cite{thiamwong2022technology,choudhury2022objectively}.
Physical activity is an integral component of a healthy lifestyle, particularly for older adults. Extensive research has highlighted the importance of physical activity for this population over a prolonged period of time \cite{seguin2002growing,nelson2007physical,evans1999exercise}. It is essential to promote and support the integration of regular physical activity in older adults. Federal physical activity guidelines recommend that older adults engage in a variety of physical activities, including balance, aerobic, and muscle-strengthening exercises \cite{piercy2018physical}.
Regular physical activity has been shown to reduce the risk of chronic diseases in older adults, such as heart disease, stroke, diabetes, and certain forms of cancer~\cite{chakravarthy2002obligation}. Additionally, physical activity can help manage pre-existing conditions like high blood pressure, arthritis, and osteoporosis~\cite{izquierdo2021international}. The World Health Organization recommends that older adults engage in at least 150 minutes of moderate or 75 minutes of vigorous physical activity per week to maintain their health and well-being \cite{world2010global}.

Precisely evaluating physical activity levels among older adults is crucial to identify individuals who require interventions to improve their activity levels and prevent functional decline. Wearable devices that use accelerometers provide a vital tool for accurately tracking physical activity. Unlike self-reported data, accelerometer data offers an objective measurement of physical activity levels that is more accurate and reliable. By recording continuous measurements of physical activity levels over several days or weeks, accelerometers provide a more comprehensive view of an individual's physical activity patterns, including the frequency, intensity, and duration of physical activity. These devices produce quantitative measures of physical activity such as step count, moderate-to-vigorous physical activity (MVPA), and overall physical activity volume, which can be used to establish goals and monitor progress over time. Standardized methods for measuring physical activity levels allow for comparisons between individuals and populations, making it easier to identify those at risk of physical inactivity and to develop targeted interventions. The small, wearable design of accelerometer devices makes them comfortable and practical for older adults to use throughout the day. However, accelerometer data can be complex and require substantial data processing to be usable, while the quality of the data can be affected by various factors such as device placement, calibration, and wear compliance. Furthermore, statistical analysis of accelerometer data can be challenging, requiring the analysis of multivariate and longitudinal data, as well as handling missing data and non-normal distributions.

The accelerometer data for physical activity assessment are characterized by significant noise and a large number of time points, which renders them difficult to interpret. While averaged curves across individuals have been utilized to visualize overall patterns, the high uncertainty associated with these curves suggests their limited utility in practice~\cite{HELWIG20163216}. 
Various statistical methods have been explored to analyze accelerometer data. 
  More objective methods are direct observation by an independent observer    and real-time recording by devices including pedometers \cite{crouter2004validity,karabulut2005comparison}, heart-rate monitors \cite{crouter2004accuracy, brage2005reliability},   armbands \cite{andre2006development}, and accelerometers \cite{grant2006validation,vanhelst2012comparison,bonomi2010estimation}. Pedometers  are  cost-efficient and convenient \cite{adamo2009comparison} although they do not measure the duration  or intensity of PA. A recent review article discussed the statistical methods for analyzing the accelerometer data~\cite{zhang2019review}. The functional data analysis framework has been used for the  minute-by-minute temporal pattern analysis of physical activity~\cite{schrack2014assessing, goldsmith2015generalized, li2017joint}. Nevertheless, there is still a gap in the statistical approach that takes into account the ANOVA-type decomposition when comparing accelerometer data across multiple groups.
  
To address these challenges, we propose to use a mixed-effects smoothing spline analysis of variance (SSANOVA) framework~\cite{gu2013smoothing, wahba1990spline, wang2011smoothing,gu2005optimal} to analyze the accelerometer data.
By utilizing the smoothing spline ANOVA model, raw accelerometer data can be processed through smoothing functions to minimize noise and enhance data quality. Additionally, this model can facilitate the identification of intricate associations between physical activity levels and variables such as age, gender, and health status. Due to its flexibility and capability to handle complex and non-linear relationships, the smoothing spline ANOVA model is a potential statistical tool that can provide precise and dependable estimates of physical activity levels and their links to health outcomes.
The smoothing spline model and its applications have been recently reviewed, and the relevant literature can be accessed through Zhang's work~\cite{zhang2018smoothing} and the references cited therein.
The SSANOVA model has been implemented in various studies across diverse fields. For instance, Davidson et al.~\cite{davidson2006comparing} employed SSANOVA to compare whole tongue contours obtained via ultrasound imaging. Meanwhile, Luo and Wahba~\cite{luo1998spatial} utilized the model to generate smoothing spline estimates for temperature data collected at multiple time points, taking into account the effects of year and location. Helwig and colleagues~\cite{helwig2015analyzing} applied the SSANOVA framework to estimate spatiotemporal trends and assess uncertainty using massive samples of social media data. Additionally, the cyclic biomechanical data analysis was conducted using the SSANOVA model~\cite{HELWIG20163216}.

In our study, we collected accelerometer data for the physical activity assessment of 121 individuals over the age of 60 for a period of 7 consecutive days, utilizing wearable activity trackers~\cite{thiamwong2021technology}. We employed ActiGraph accelerometers (ActiGraph GT9X Link wireless, ActiGraph LLC.) to collect the data, offering an impartial evaluation of physical activity that is not dependent on self-reporting.
Our aim was to use a mixed-effects smoothing spline analysis of variance (SSANOVA) framework to analyze accelerometer data, gaining a better understanding of older adults' physical activity levels and promoting the importance of implementing balance-improving interventions and cognitive restructuring to encourage physical activity among older adults.

The structure of this paper is as follows:  Section \ref{ch:review} outlines the smoothing spline analysis of variance model.  Section~\ref{ch:anova} presents the descriptive statistics of accelerometer data and demographics of older adults. In this section, we also apply mixed-effects SSANOVA models to estimate the smooth functions from the noisy measurements obtained from the Actigraph device.   Section \ref{ch:con} provides a conclusion and discusses future directions.

\section{Smoothing Spline Analysis of Variance (SSANOVA)}\label{ch:review}
We propose a mixed-effects SSANOVA model of the form 
\begin{equation}\label{eq:ssmodel}
y_i=\eta(\mathbf{x}_i)+b_{s_i}+\epsilon_i, 
\end{equation} 
where $y_i\in \mathbb{R}$ is the recorded physical activity at time point $i$, $\mathbf{x}_{i} = (t_i, g_i)$ is the fixed effect predictor vector corresponding to the $i$-th data point with $t_i$ denoting the time point  and $g_i$ denoting the groups, $s_i$ is the subject indicator, $b_{s_i}$ is the subject random effect such that $b_{s_i}\overset{iid}{\sim}N(0, \sigma^2_{s_i})$, and $\epsilon_i\overset{iid}{\sim}N(0, \sigma^2_{\epsilon} )$ is random error that is independent of the random effects, for $i = 1,\ldots, T$ where $T$ is the total number of data points after vectorizing the data. 

The rationale behind the SSANOVA is that the smooth function $\eta(\mathbf{x}_i) = \eta(t_i,g_i)$ is decomposed into components associated with time $t_i$ and group member $g_i$ as 
\begin{equation}
    \eta(t_i,g_i)=\eta_0+\eta_1(t_i )+\eta_2( g_i)+\eta_{12}(t_i,g_i),
\end{equation}
where $\eta_0$ is a constant function, $\eta_1$ is the main effect of time , $\eta_2$ is
the main effect of the group, and $\eta_{12}$ is the time-group interaction effect. The SSANOVA model makes it possible to examine the physical activity difference between two groups
\begin{equation}
    \label{diff}
    \delta(t|g,g_{*}) = \eta(t,g) - \eta(t,g_{*})
\end{equation}
for different groups $g$ and $g_{*}$. The functional differences  $\delta(t|g,g_{*})$ quantify the differences between the physical activity  of groups $g$ and $g_{*}$ across the entire study period, and can be used to identify time regions where functional differences exist between the groups, instead
of the traditional pointwise comparisons for the statistical analysis of accelerometer data.

 The function $\eta(t_i,g_i)$ is estimated by minimizing the penalized least squares \cite{gu2013smoothing},
\begin{equation}
  \frac{1}{n}\sum_{i=1}^{n}(y_i-(\eta_0+\eta_1(t_i )+\eta_2( x_i)+\eta_3(t_i,x_i)))^2+\lambda\int_0^1(\eta''(t_i,x_i))^2dx  \label{eq:penalty}
\end{equation} 
where the smoothness penalty term $\int_0^1(\eta''(t_i,x_i))^2dx$ is nonnegative definite. The model’s optimal smoothing parameters $\lambda$ were estimated using the generalized cross-validation method~\cite{craven1978smoothing}.

We can construct the Bayesian confidence intervals (CIs) for the estimated function $\hat{\eta}$ and $\hat{\delta}$ via the Bayesian interpretation of a smoothing spline, where the significant functional difference can be determined and  uncertainty of the estimation can be quantified~\cite{wahba1983bayesian,gu1993smoothing}. Thus we utilize $\hat{\delta}$ and CIs to test for differences between the groups and identify the time regions  of statistically significant differences in physical activities. 
Time points or region where the confidence interval of $\hat{\delta}$ does not include zero are considered to demonstrate significant mean differences between the groups. SSANOVA models hence provide an accurate and powerful tool to identify the time regions of statistically significant differences in physical activity between groups.

\section{Application of Smoothing Spline ANOVA to Accelerometer Data}\label{ch:anova}

\subsection{Accelerometer Data}
We summarize the demographic information, fall-related variables and physical activity measurement vector magnitude (VM) for the 121 participants.
The participants were divided into four groups based on their physical abilities and fall risk appraisal: \textit{rational}, \textit{irrational}, \textit{congruent}, and \textit{incongruent}~\cite{thiamwong2021technology}.
The rational group consists of individuals with normal balance and a normal perceived risk of falls. 
The irrational group has normal balance ability, but with higher perceived risk of falls. Both the congruent and incongruent groups have poor balance, but congruent group exhibits a low perceived risk of falls, while the incongruent group presents a high perceived risk of falls 

\begin{figure}
  \centering
            \includegraphics[width=\textwidth]{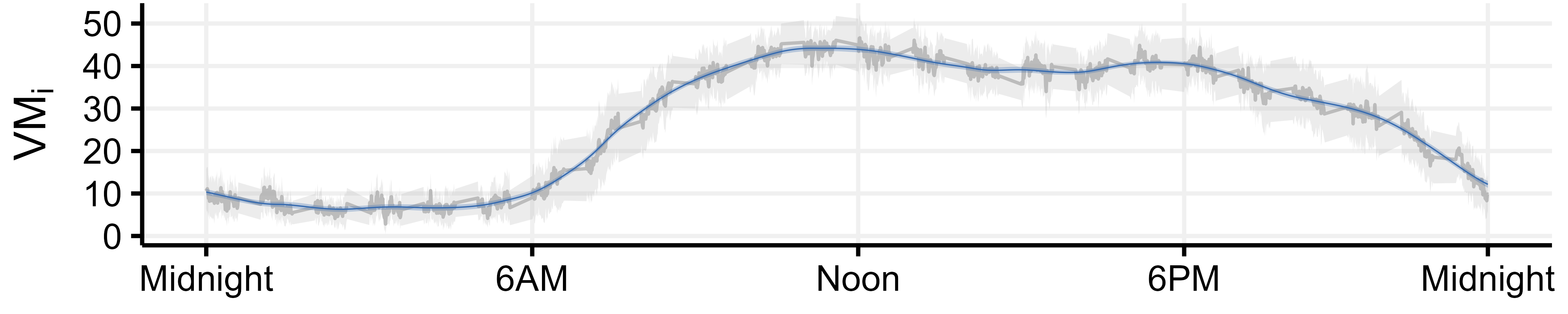}
        \caption
        {SSANOVA curves (blue) and the sample average (grey) of daily physical activity for congruent group participants.} 
        \label{fig:curves}
    \end{figure}

Table \ref{tab:summary table} reports summary statistics of  age, gender, history of falling, and history of injurious falls, VM, time by fall risk groups, where the VM records were aggregated to one-minute resolution.
Figure~\ref{fig:curves} depicts the daily physical activity represented as vector magnitude (VM) for the congruent group participants. 
The sample average over time (Figure~\ref{fig:curves}, grey) shows considerable fluctuation across individuals, while SSANOVA curve  (Figure~\ref{fig:curves}, blue) more clearly visualize overall daily physical activity patterns.

\begin{table}[!htbp]
 \scriptsize\centering
  \caption{Summary of Response Variable and Independent Variables}
 
    \begin{tabular}{lrrrrrrrrrrr}\hline
    Variable & \multicolumn{1}{l}{Mean} & \multicolumn{1}{l}{Std} & \multicolumn{1}{l}{Min} & \multicolumn{1}{l}{Median} & \multicolumn{1}{l}{Max} &       & \multicolumn{1}{l}{Mean} & \multicolumn{1}{l}{Std} & \multicolumn{1}{l}{Min} & \multicolumn{1}{l}{Median} & \multicolumn{1}{l}{Max} \\\hline
    \multicolumn{6}{c}{Congruent Group ($n=19$)} & \multicolumn{6}{c}{Irrational Group ($n=22$)} \\\cline{2-6}  \cline{8-12}  VM    & 952.0 & 1789.4 & 0.0   & 0.0   & 32439.0 &       & 971.9 & 1850.5 & 0.0   & 0.0   & 32223.0 \\
    Time  & 1177.9 & 691.8 & 0.0   & 1157.0 & 2359.0 &       & 1180.9 & 691.6 & 0.0   & 1156.0 & 2359.0 \\
    Age   & 78.3  & 7.1   & 67.0  & 79.0  & 93.0  &       & 74.0  & 7.4   & 60.0  & 75.0  & 90.0 \\
    Female & $80\%$   &       &       &       &       &       & $90\%$   &       &       &       &  \\
    History of Falls & $35.0\%$   &       &       &       &       &       & $22.7\%$   &       &       &       &  \\
    History of Injurious Falls & $8.4\%$   &       &       &       &       &       & $6.3\%$   &       &       &       &  \\
   \hline
    \multicolumn{6}{c}{Incongruent Group ($n=23$)} & \multicolumn{6}{c}{Rational Group ($n=57$)} \\\cline{2-6}  \cline{8-12}
    VM    & 887.9 & 1797.2 & 0.0   & 0.0   & 37425.0 &       & 907.6 & 1956.8 & 0.0   & 0.0   & 31737.0 \\
    Time  & 1177.1 & 691.8 & 0.0   & 1200.0 & 2359.0 &       & 1178.5 & 692.1 & 0.0   & 1157.0 & 2359.0 \\
    Age   & 77.6  & 6.3   & 68.0  & 76.0  & 94.0  &       & 72.2  & 6.3   & 60.0  & 71.0  & 96.0 \\
    Female & $60\%$   &       &       &       &       &       & $80\%$   &       &       &       &  \\
    History of Falls & $36.9\%$   &       &       &       &       &       & $44.2\%$   &       &       &       &  \\
    History of Injurious Falls & $8.5\%$   &       &       &       &       &       & $9.9\%$   &       &       &       &  \\
     \hline 
    \end{tabular}%
 \label{tab:summary table}%
\end{table}%

\subsection{SSANOVA by Fall Risk Groups for Daily Physical Activity}

 Our objective was to estimate the functional trends of physical activity for these four groups and investigate the variability of the estimated trends~\cite{thiamwong2020assessing,choudhury2023associations}.
Using the mixed-effects smoothing spline ANOVA model proposed in Eq.~\ref{eq:ssmodel}, we fit the log-transformed VM on time $t$ and fall risk group $g$. The SSANOVA models are fit using 
 {\em gss} package~\cite{gu2014smoothing} in R 
 environment~\cite{rlang}. 
 The cubic smoothing spline was used for daily (in minutes) $t \in (0,1440)$, and fall risk group $g$ (nominal variable: rational, irrational, congruent, and incongruent). 
 We estimate the variances of the random subject effects and the random errors (i.e., $\sigma_{b_i}$ and $\sigma^2_{\epsilon}$) from the data using a restricted maximum likelihood approach\cite{gu2013smoothing,helwig2016efficient}. 
 The optimal values of smoothing parameters in the models were chosen using the generalized cross-validation method.

\begin{figure}[ht]
\vspace{-.1in}
\includegraphics[scale=0.75]{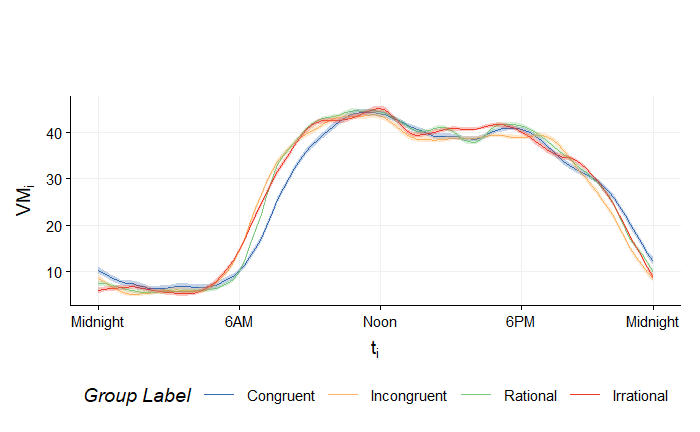}
\caption{SSANOVA curves and $95\%$ CI for VM over 24 hours by fall risk groups.}
\vspace{-.1in}
\label{fig:VMTimeH}
\end{figure}

Figure \ref{fig:VMTimeH} plots the estimated functions of daily VM along with $95\%$ CIs by fall risk group.
The corresponding model has total $R^2 =62.2\%$ and estimated error variance $\hat{\sigma}_{\epsilon}= 11.51$. 
Compared to the sample average, the SSANOVA model takes into account both the within and between subject variability in the data, leading to significant reductions in standard errors and thus narrower CIs.
The $95\%$ Bayesian confidence interval indicates that the rational group had the highest physical activity (PA) level while the congruent group had significantly lower PA levels than the other three groups. Additionally, there was a noticeable periodic trend in PA levels observed in all study samples regardless of their fall risk groups, with inactivity during the night (12-5am), an increasing trend in the morning (5am-noon), a peak at noon, a moderate decrease in the afternoon (noon-6pm), and a significant decline in the evening.

\subsection{SSANOVA for the History of Falls and Injurious Falls}

\begin{figure}
    \centering
    \begin{subfigure}[b]{0.48\textwidth}
        \includegraphics[width=\textwidth]{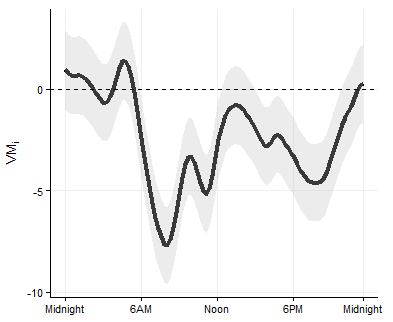}
        \caption{Congruent}
    \end{subfigure}
    \hfill
    \begin{subfigure}[b]{0.48\textwidth}
        \includegraphics[width=\textwidth]{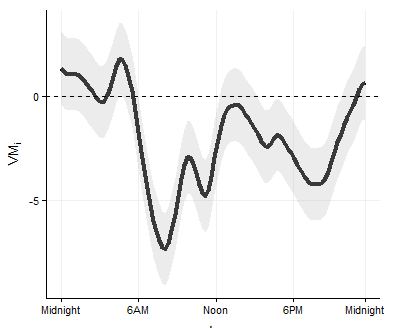}
        \caption{Incongruent}
    \end{subfigure}

    \begin{subfigure}[b]{0.48\textwidth}
        \includegraphics[width=\textwidth]{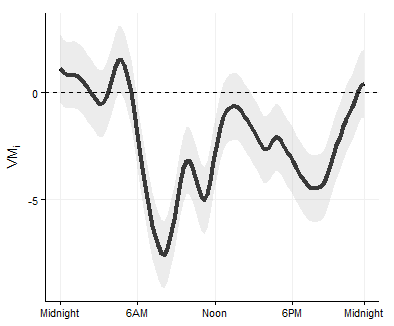}
        \caption{Rational}
    \end{subfigure}
    \hfill
    \begin{subfigure}[b]{0.48\textwidth}
        \includegraphics[width=\textwidth]{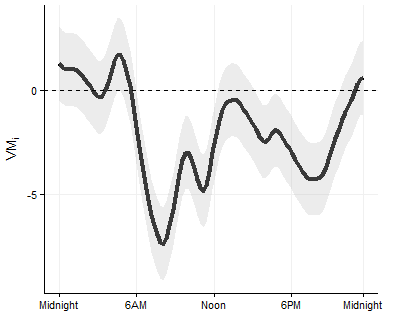}
        \caption{Irrational}
    \end{subfigure}

    \caption{Physical activity intensity difference $\hat{\delta}_{\text{VM}}$ and $95\%$ CIs between groups with and without fall history.}\label{fig:mainfig}
\end{figure}

\begin{figure}
    \centering
    \begin{subfigure}[b]{0.48\textwidth}
        \includegraphics[width=\textwidth]{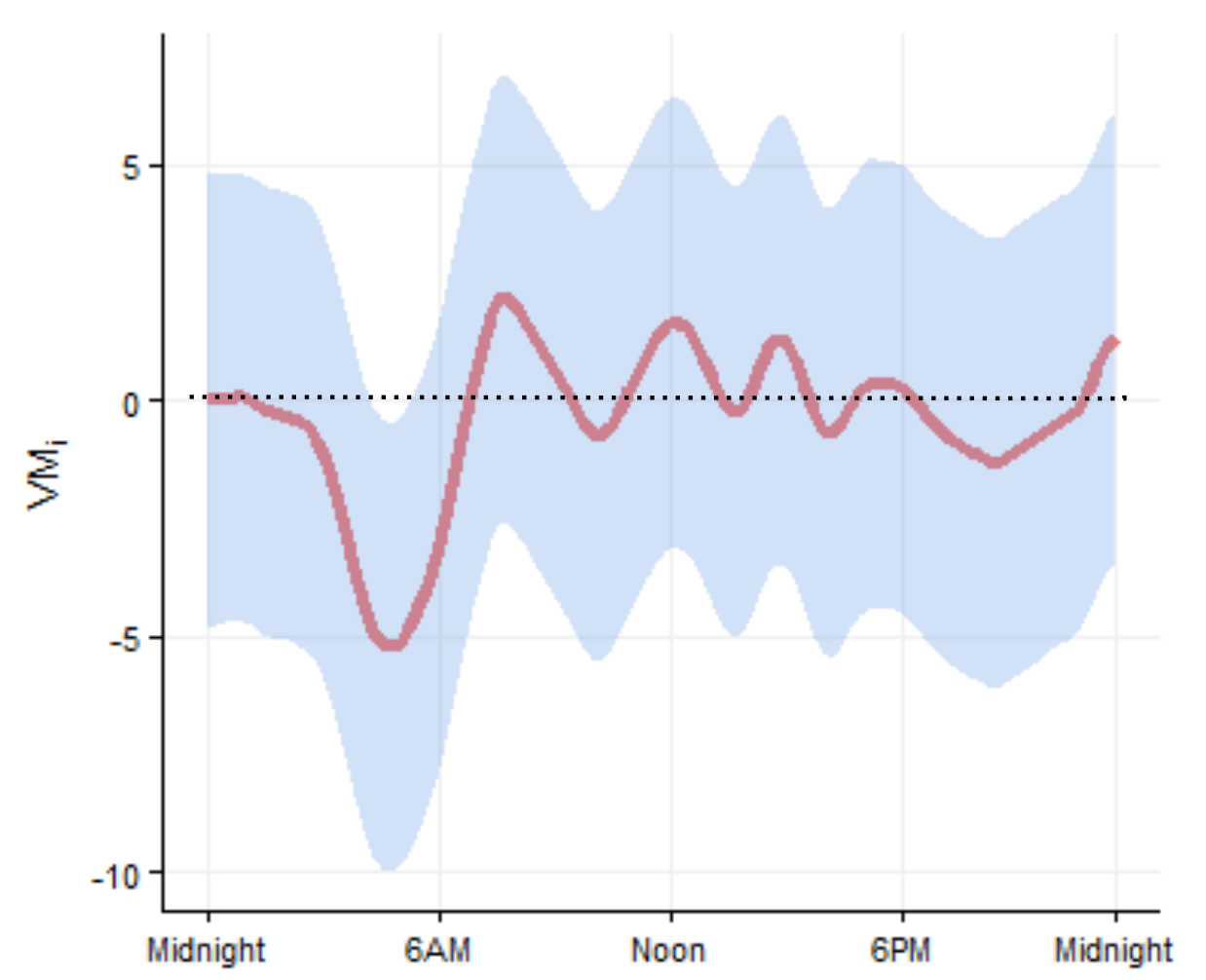}
        \caption{Congruent}
    \end{subfigure}
    \hfill
    \begin{subfigure}[b]{0.48\textwidth}
        \includegraphics[width=\textwidth]{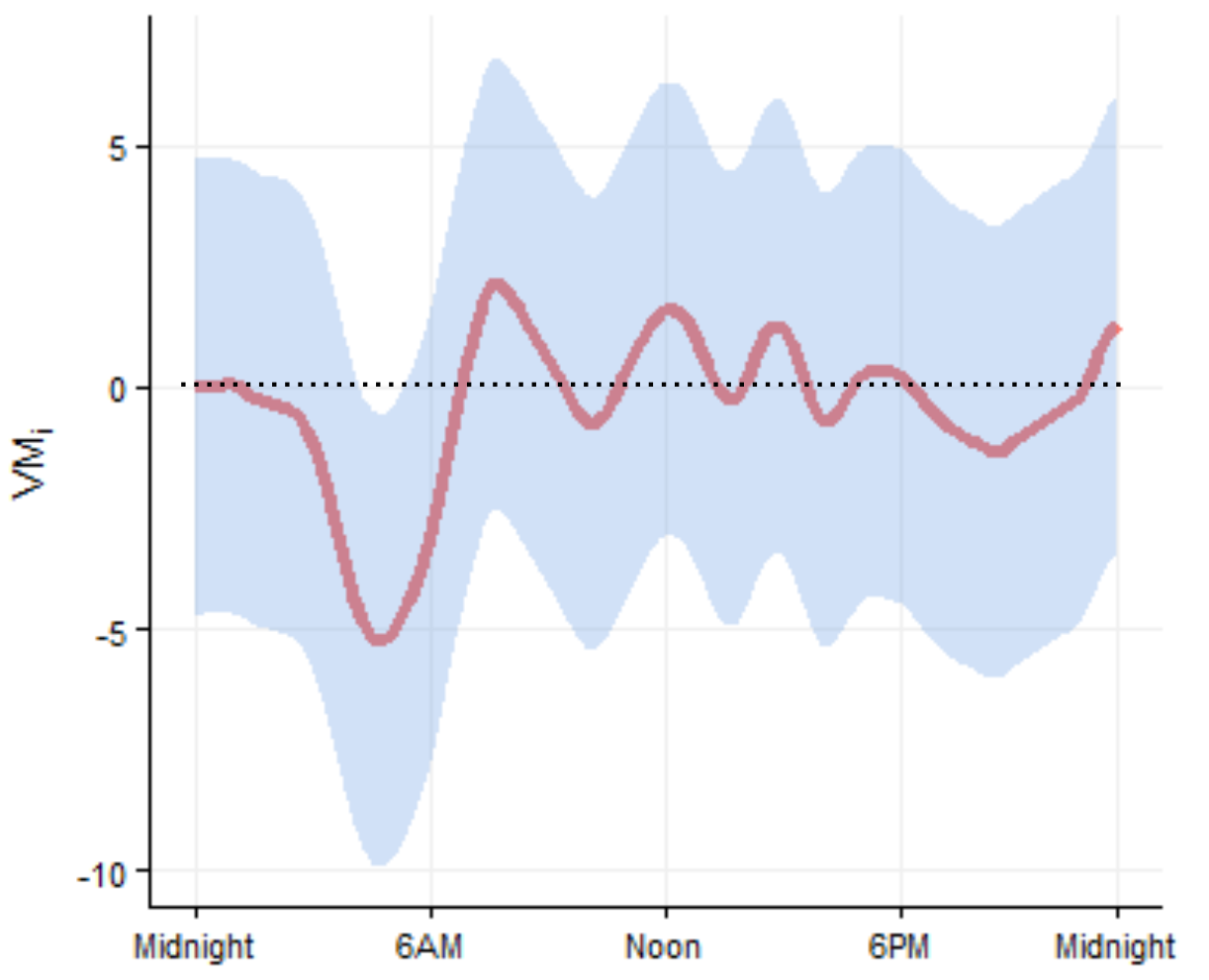}
        \caption{Incongruent}
    \end{subfigure}

    \begin{subfigure}[b]{0.48\textwidth}
        \includegraphics[width=\textwidth]{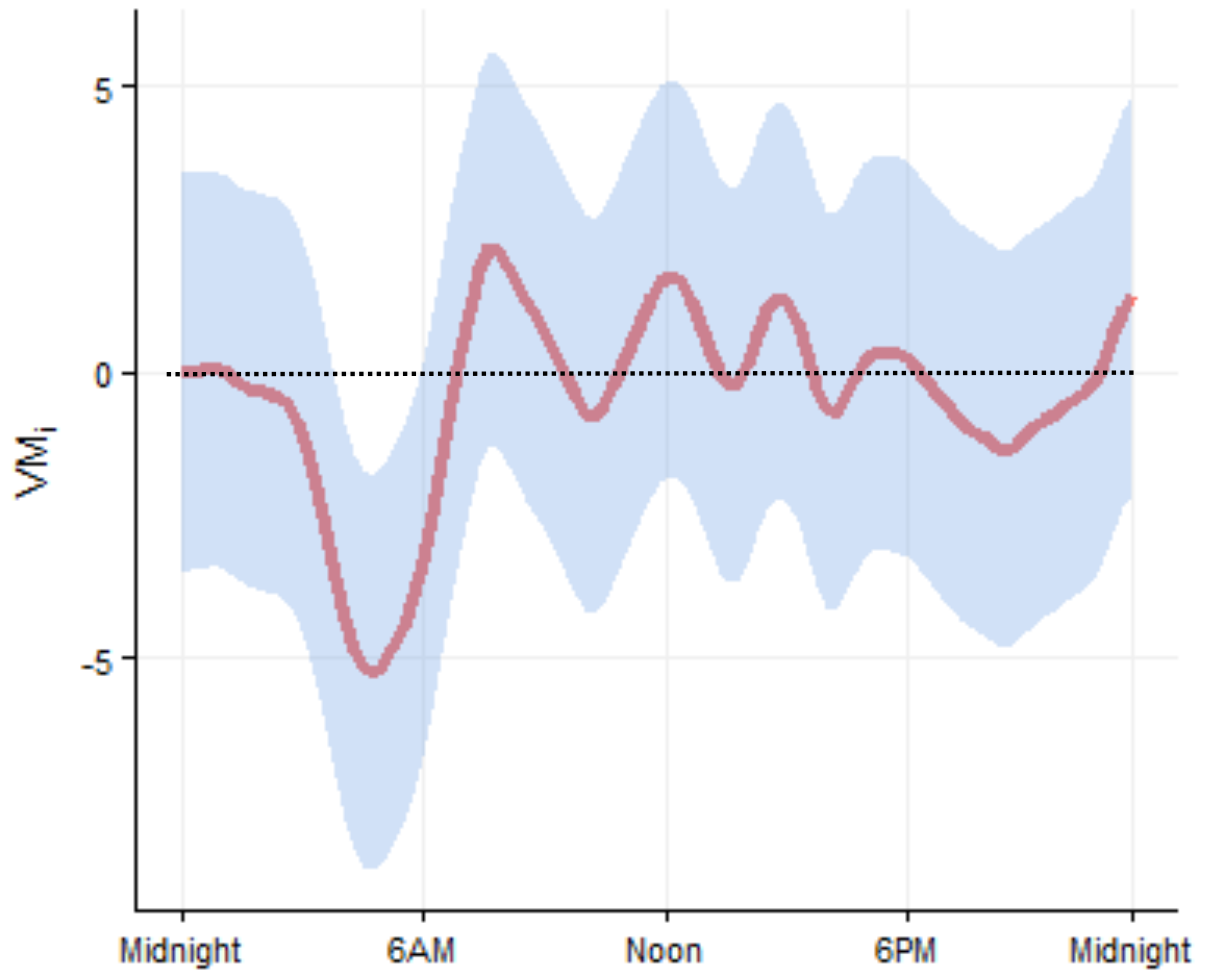}
        \caption{Rational}
    \end{subfigure}
    \hfill
    \begin{subfigure}[b]{0.48\textwidth}
        \includegraphics[width=\textwidth]{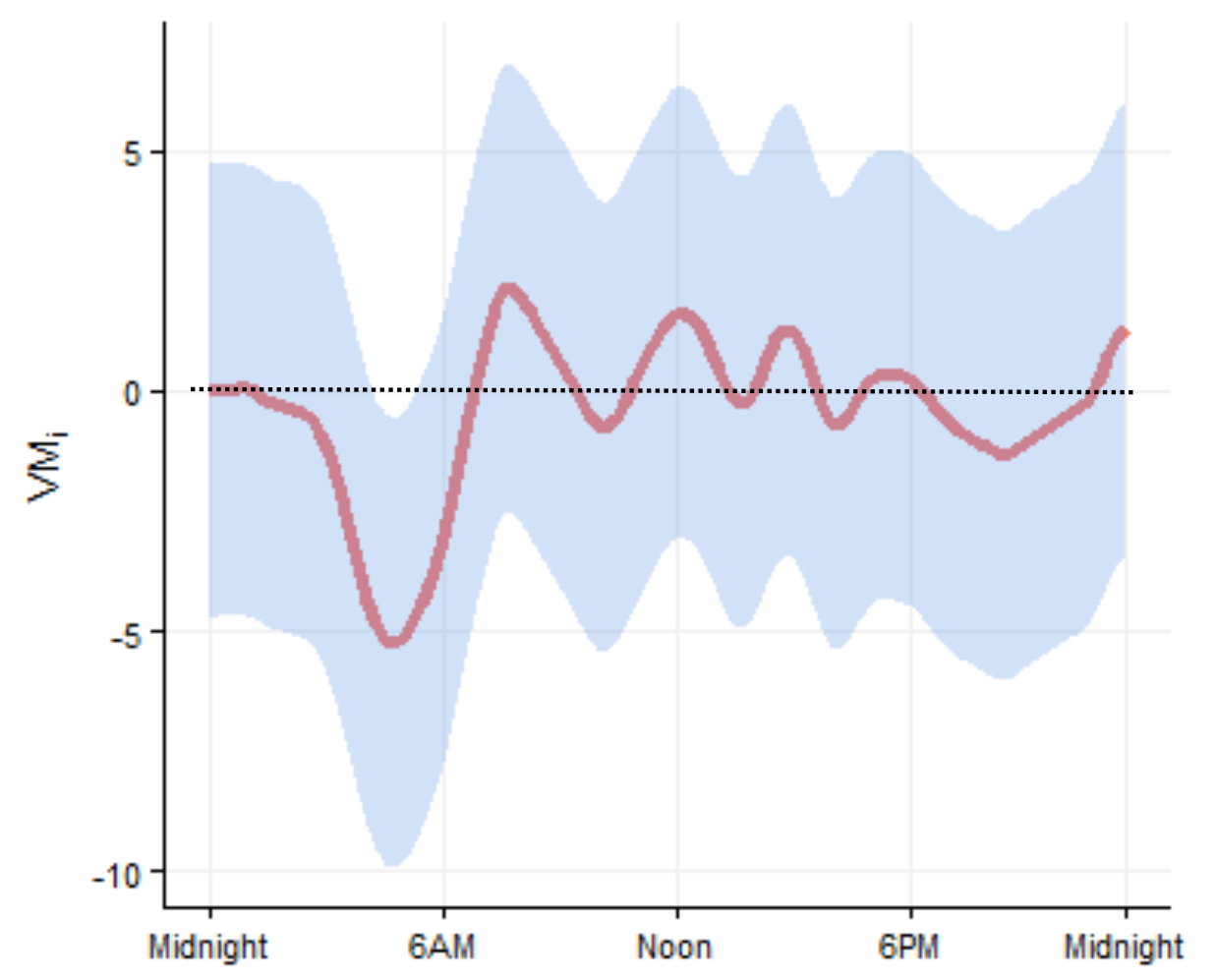}
        \caption{Irrational}
    \end{subfigure}

    \caption{Physical activity intensity difference $\hat{\delta}_{\text{VM}}$ and $95\%$ CIs between groups with and without history of injurious falls .}\label{fig:mainfig1}
\end{figure}

If we take the \textit{history of falls} and \textit{history of injurious falls} into account as additive functions, the SSANOVA model becomes
\begin{align*}
{\log(\text{VM})}_i&=\eta_1(t_i)+\eta_2(g_i) +\eta_{12}(t_i,g_i)\\ 
&+\eta_3(\text{Falls}_i)+\eta_4(\text{Fall-related Injury}_i)+b_{s_i}+\epsilon_i, 
\end{align*}
where the groups with and without history of falls and groups with and without history of injurious falls are compared via $\eta_3(\cdot)$ and $\eta_4(\cdot)$, respectively, using dummy variables as inputs.
Figure~\ref{fig:mainfig} provides the estimated SSANOVA curves for the difference in physical activity intensity $\hat{\delta}_{\text{VM}}$ across 24 hours between the groups with and without history of falls. The $95\%$ CIs for $\hat{\delta}_{\text{VM}}$ is also estimated (grey bands in Figure~\ref{fig:mainfig}).
The results indicate that the group of participants with history of falls have significantly higher physical activity intensity during the daily activity (6 am - 10 pm) compared to those without history of falls, regardless of the fall risk groups, since the CIs in Figure~\ref{fig:mainfig} are below 0 reference line from 6 am to 10 pm. 

Similarly, Figure~\ref{fig:mainfig1} provides the estimated SSANOVA curves for the difference in physical activity intensity $\hat{\delta}_{\text{VM}}$ and $95\%$ CIs (blue bands) across 24 hours between the groups with and without history of injurious falls. For rational group, participants with history of injurious falls have significantly higher physical activity intensity during early morning (3 - 6 am) compared to those only with history of falls but no injurious falls. For congruent, incongruent and irrational groups, similar pattern observed with much narrow significant time region.

\section{Conclusion} \label{ch:con}
The paper utilizes SSANOVA models on accelerometer data to estimate physical activity patterns in community-dwelling older adults and differences among fall risk groups. The analysis reveals distinct patterns of activity levels throughout the day and across days, with a noticeable periodic trend in physical activity, including inactivity during night (12-5am), an increasing trend in the morning (5am-noon), a peak at noon, a moderate decrease in the afternoon (noon-6pm), and a significant decline in the evening.
The study also finds that the group-specific physical activity patterns within a day. In particular, the congruent group has significantly lower physical activity levels than the other three groups, while the rational group has the highest physical activity levels, based on $95\%$ Bayesian confidence intervals. 
These results offer important insights into the daily and longitudinal activity patterns of different groups that can be used to tailor interventions based on the fall risk groups, balance performance, fear of falling levels and time preference. 
Moreover, the results can inform interventions to enhance physical activity levels and improve overall health outcomes for older adults.

Additionally, this study investigates the daily physical activity levels and pattern between history of falls, and injurious falls. The outcomes reveal noticeable variations in physical activity levels between groups with history of falls or not.
Additionally, individuals with a history of injurious falls exhibit higher levels of physical activity compared to those without such injuries, especially in the early morning (12 - 6 am).

As SSANOVA models provide a powerful framework for analyzing cyclic or periodic data, future research can explore their potential in analyzing accelerometer data to identify more complex patterns of physical activity.
Overall, the use of SSANOVA models in accelerometer data analysis holds great potential for advancing our understanding of physical activity patterns and their association with health outcomes in older adults, which can inform the development of targeted interventions to improve their overall health and well-being.

One potential future research direction using SSANOVA models on analyzing accelerometer data could be to focus on mobile health and smart intervention using wearable devices. This could involve using SSANOVA models to better understand patterns of physical activity and sedentary behavior in real-time, as well as how these patterns relate to health outcomes such as obesity, diabetes, and cardiovascular disease. Additionally, SSANOVA models could be used to develop personalized intervention strategies based on an individual's unique activity patterns, with the goal of promoting healthy behavior change and preventing disease. Another potential direction could be to incorporate additional sensors, such as heart rate monitors or GPS trackers, to better understand the context in which physical activity occurs and how it relates to health outcomes. Overall, the use of SSANOVA models in mobile health and intervention research has the potential to provide valuable insights into how technology can be leveraged to improve health outcomes and promote healthy behavior change.

\bibliographystyle{ieeetr}
\bibliography{FDAref}
\end{document}